# A complete fs-laser-ablation route to miniaturized single-crystal PMN-PT piezoelectric actuators

*Menotti Markovic[1], Lucia Oberndorfer[1], Tobias M. Krieger[2], Ievgen Brytavskyi[2], Barbara Lehner[2], Julia Freund[3], Vishnu Prakash Karunakaran[1,2], Matthias Domke[1], Dorian Gangloff[4], Christian Schimpf[4], Peter Michler[5], Simone Luca Portalupi[5], Michael Jetter[5], Rinaldo Trotta[6], Javier Martín-Sánchez[7], Armando Rastelli[2], Fadi Dohnal[1], Sandra Stroj[1]*

*[1] Research Center for Microtechnology, Vorarlberg University of Applied Sciences, Hochschulstraße 1, A-6850 Dornbirn, Austria*
*[2] Institute of Semiconductor and Solid State Physics, Johannes Kepler University, Altenbergerstraße 69, A-4040 Linz, Austria*
*[3]Department of Theoretical Physics, University of Innsbruck, Technikerstraße 21A A-6020 Innsbruck*
*[4] Cavendish Laboratory, University of Cambridge, J.J. Thomson Avenue, Cambridge, CB3 0HE, UK*
*[5] Institute of Semiconductor Optics and Functional Interfaces, University of Stuttgart, Stuttgart, Germany*
*[6]Department of Physics, Sapienza University of Rome, Piazzale A. Moro 5, 00185 Rome, Italy*
*[7] Department of Physics, University of Oviedo, C/ Federico García Lorca nº 18, 33007, Oviedo, Spain*

**Abstract:** This article presents a novel fabrication route for miniaturized piezoelectric actuators that relies exclusively on processes based on femtosecond (fs) laser ablation. Previous work has already demonstrated that fs-lasers are uniquely suited for the fabrication of piezoelectric actuators based on PMN-PT, which are required for multiaxial strain-tuning of quantum dots (QDs) to enable, e.g. the generation of highly entangled photon pairs. Building on these foundations, the present work advances actuator performance and capabilities by introducing a local thinning strategy. This approach allows the realization of smaller devices, which in turn enables lower operating voltages, while simultaneously offering the possibility of integrating multiple quantum light sources on a single chip. The article provides a detailed description of the full fabrication chain, entirely based on fs-laser processing steps, from substrate thinning to metal layer structuring and final device definition. A particular focus is placed on the final cutting process, where the implementation of a third-harmonic ultraviolet (UV) fs-laser wavelength significantly improves edge quality and shape definition compared to the second harmonic (SH) wavelength used in previous work. The device fabricated through the combination of local thinning and UV-based cutting promises not only to enhance the efficiency of strain transfer but also to ensure the mechanical stability required for practical applications. These results establish fs-laser-based fabrication as a versatile and scalable method for next-generation piezoelectric actuators, paving the way for advanced strain-engineering approaches in semiconductor quantum optics and integrated quantum photonics.

## 1. Introduction

Piezoelectric materials play a crucial role in Micro-Electro-Mechanical Systems (MEMS) due to their unique ability to convert mechanical into electrical energy and vice versa. This makes them well suited for sensors and actuators, offering a large dynamic range, low power consumption, and minimal hysteresis [1]. Among the various piezoelectric materials, lead magnesium niobate-lead titanate (PMN-PT) in single crystal form stands out due to its exceptionally high piezoelectric constant ($d_{33}$) and coupling factors, as well as its high sensitivity. These characteristics facilitate its use in high-quality sensors and actuators [2] [3] [4]. Notably, PMN-PT materials maintain their effective performance even at cryogenic temperatures, making them suitable for applications in low-temperature environments [5].

Actuators based on single-crystal PMN-PT are delicate due to brittleness, small feature sizes, and thin substrates, which restricts available processing options. For reproducible, reliable device function and acceptable yield, micrometer-scale resolution, arbitrary geometries, low surface roughness, and sharp edges are required. Various structuring methods have been explored for PMN-PT, including wet etching [6], dry etching [7] [8], reactive ion etching [9] or ion milling [3], [10]. Each process has its advantages, but also limitations such as long processing times, limited geometric possibilities or insufficient processing quality. Laser ablation in general addresses several of these constraints. It provides a flexible direct-writing method with competitive processing times and the potential for high-quality results [7], [11], [12], [13]. Micromachining with ultrashort laser pulses in particular has established itself as a reliable technique for precise material removal while minimizing thermal load and collateral damage to the surrounding regions [14]. This enables highly controlled, high-quality processing and has also proven effective for machining particularly hard and brittle materials such as PMN-PT [15], [16], [17].

Building on these capabilities, our previous work demonstrated the potential of femtosecond laser processing for PMN-PT devices relevant to semiconductor-based quantum optics [18]. Actuators of various geometries were realized by laser cutting using the second harmonic of a femtosecond source at 523 nm, and the electrical contacts were patterned by selective ablation of a gold layer with the same laser. The resulting devices featured a total thickness of 300 µm and a side length of 5 mm, with a minimum spacing between actuator arms of 30 µm, limited by laser parameters and sample thickness. Having established a reliable fabrication route, we employed these actuators as strain-tuning devices for semiconductor quantum dots (QDs). Their precise controllability and strong piezoelectric response are essential for this application, as strain directly modifies the band structure and optical properties of QDs under cryogenic conditions. Controlled in-plane strain fields in QDs enable, for example, the production of polarization-entangled photons by tuning the fine-structure splitting in the biexciton–exciton cascade [19]. It has already been successfully demonstrated that our multi-axis piezo actuators enable emission energy tuning of quantum dots as well as to generate entangled photon pairs [20], [21], [22], [23].

Current efforts in quantum-emitter integration target higher device densities, driving a need for miniaturized actuators. Our new design strategy therefore introduces localized thinning of the active regions (Fig. 1). This increases the electric field at a given voltage while preserving mechanical stability through thicker surrounding areas. The actuator length change follows $\Delta L/L = d_{33}E$, where the strain depends on the piezoelectric coefficient $d_{33}$ and the electric field $E$. With $E = V/t$, (applied voltage $V$, substrate thickness $t$), the displacement can be written as $\Delta L = Ld_{33}V/t$. Hence, the response depends on electric field rather than absolute voltage, and the voltage required for a target strain scales linearly with thickness. Thinner regions therefore achieve the same strain at substantially lower voltages [24]. In addition, laser cutting within these thinned regions enables the realisation of narrower cutting widths offering the prospect of higher strain amplification. However, this new generation of actuators leads to a more complex manufacturing workflow compared to the previously published strain-tuning platforms. For the final device, it is critical that the fabrication produces clean, defect-free surfaces – without debris, cracks, or fracture points – as these affect its functionality during use.

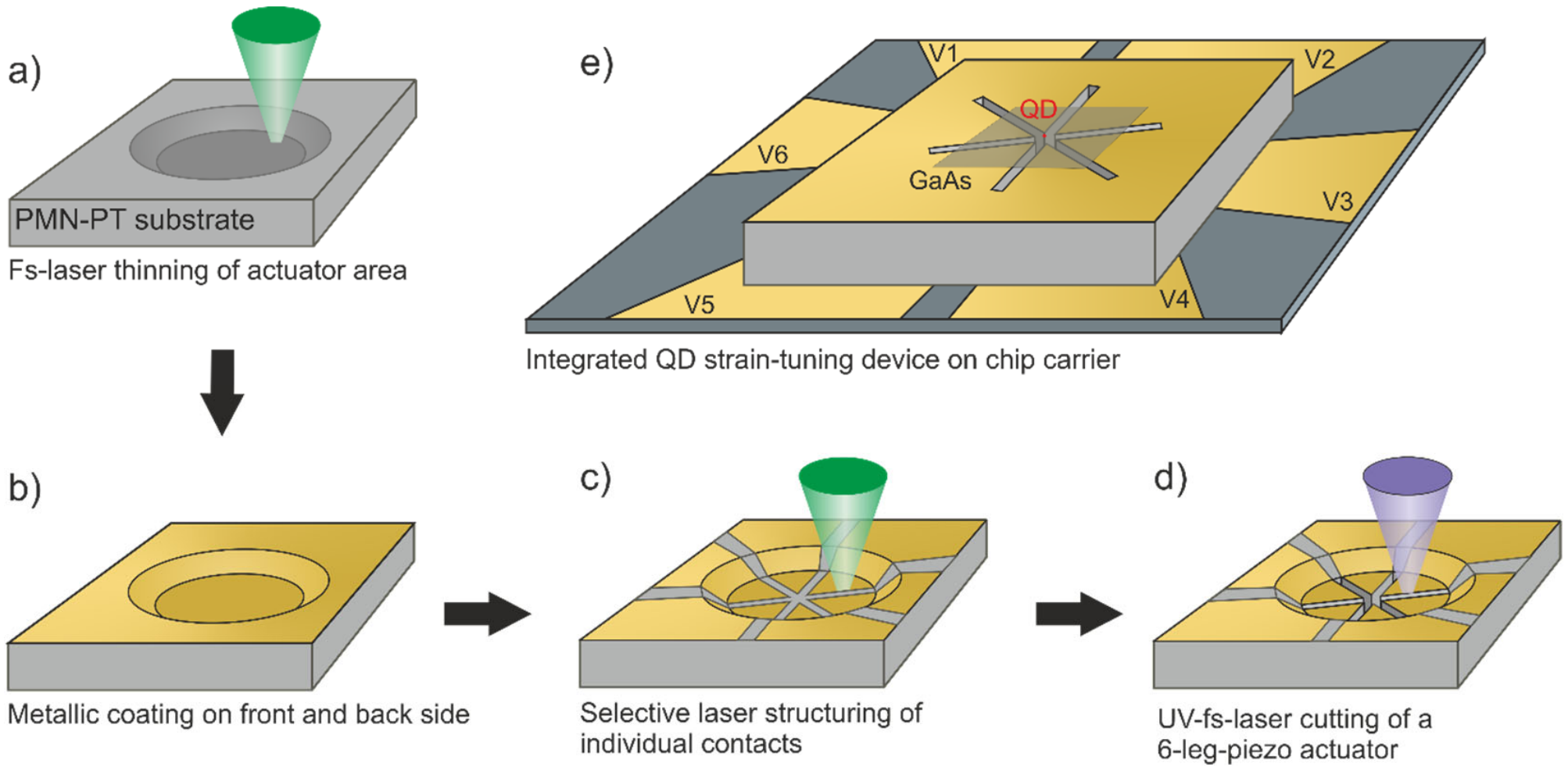

**Fig. 1** The schematic diagram shows the process workflow for realising locally thinned piezo actuators using fs-laser ablation and a sketch of a strain-tuned quantum emitter. In the first step, the 300 µm thick single-crystal material is locally thinned **(a)** using the second harmonic of the femtosecond laser. Then, thin metal layers are vapour-deposited on the front and back side **(b)**. These serve as an electrical ground on the bottom and contacts for the individual actuator arms on the top side. An intact conductive layer over the cavity edges is important to ensure reliable control of the individual actuator arms. After structuring of the front contacts by means of selective ablation **(c)**, the actuator is finally created using UV-fs-laser cutting **(d)**. **(e)** Sketch of a 6-legged actuator for multiaxial strain tuning of a QD embedded in a GaAs nanomembrane bonded on its top surface. Six independent voltages ($V_1$- $V_6$) are applied to the individual legs at the bottom of the actuator by contacting each of the legs on a chip carrier.

This work covers the development of the process steps for thinning (Fig. 1a), contact layer structuring (Fig. 1c), and UV cutting (Fig. 1d). In addition to the development of the laser processes, the influence of the metal layer thickness on the conductivity over the non-planar areas is explored (Fig. 1b). The final integration of the actuator (see Fig. 1e) is identical to previous work, with the difference that local thinning creates a space between the PMN-PT and the chip carrier in the active area. This is an improvement as the free movement of the individual arms is always guaranteed.

Compared to actuators in previous work [25], the experimental design was miniaturized in terms of lateral diameter and cutting width. It consists of two symmetrical sides with two cantilevers (arms) and a center bar, a configuration which enables the individual manipulation of strain in two regions in close vicinity to each other.

# 2. Experimental

The material used for all experiments was x2b from TRS ceramics with a thickness of 300 µm. A femtosecond laser source (Spirit$^{(R)}$ 1040-HE from Spectra Physics) with a pulse duration of 330 fs (260 fs @ SH) was used for the ablation experiments. The laser is integrated in a micromachining system "microstructVario" from 3D Micromac. For most experiments the second harmonic at a wavelength of 523 nm was used. The beam was focused and manipulated using a galvo scanner with a lens of f=170 mm focal length. The cutting experiments were performed with the third harmonic (UV) wavelength at 348 nm using a f=160 mm lens. The beam radius was determined with the method of Liu et al. [26] and was 11 µm for the 523 nm wavelength and 6 µm for the UV wavelength, respectively. A repetition rate of 100 kHz was used for all thinning and layer structuring experiments. The UV cutting experiments were performed with a repetition rate of 250 kHz. Appendix A summarizes supporting investigations, including the determination of the ablation threshold (A.1) as well as measurements of the UV-laser ablation efficiency (A.2), and Raman spectroscopy of laser-processed surfaces (A.3). These complementary studies provide the key parameters and verification needed to define and validate the processing conditions applied in this work.

The substrates were coated using vacuum evaporation with an e-gun. The machine used was a Balzers BAK 550 with ESQ 110 e-gun. The source material was held in a water-cooled graphite crucible. The profile measurements of cavities were done with a Keyence VK-X250 Laser scanning microscope.

# 3. Results

## 3.1 Generation of cavities with freely adjustable side wall angles

As the first step of the actuator fabrication process, the PMN-PT substrate with a thickness of 300 µm must be thinned down to a residual thickness of 100 microns within the area of a circle (see Fig. 1a). Thinning is achieved by scanning the laser along parallel lines separated by 10 µm. This line spacing is chosen to match the effective pulse-to-pulse distance of 10 µm - set by a scan speed of 1 m/s at a 100 kHz repetition rate - and it approximately corresponds to the laser beam radius at focus. Aligning the spatial pulse spacing with the beam radius has been shown to be ideal for homogeneous material removal. It ensures sufficient areal coverage with minimal pulse overlap, thereby limiting heat accumulation. In addition, the scanning pattern is rotated by a random value after each repetition. This ensures that line intersections are statistically distributed over the surface. For a specific set of laser and structuring parameters, the ablation depth per repetition was determined. For this, the depth of a cavity was measured after a repetition number of N=200 using an optical microscope which resulted in a value of 150 µm. Based on this measurement there is a focus shift of 150 µm/N=0.75 µm applied after each repetition for all following experiments. This focus shift leads to a constant fluence value on the substrate surface during the machining process and results in a simple linear relation between ablation depth and number of repetitions which is shown in Fig. 2a.

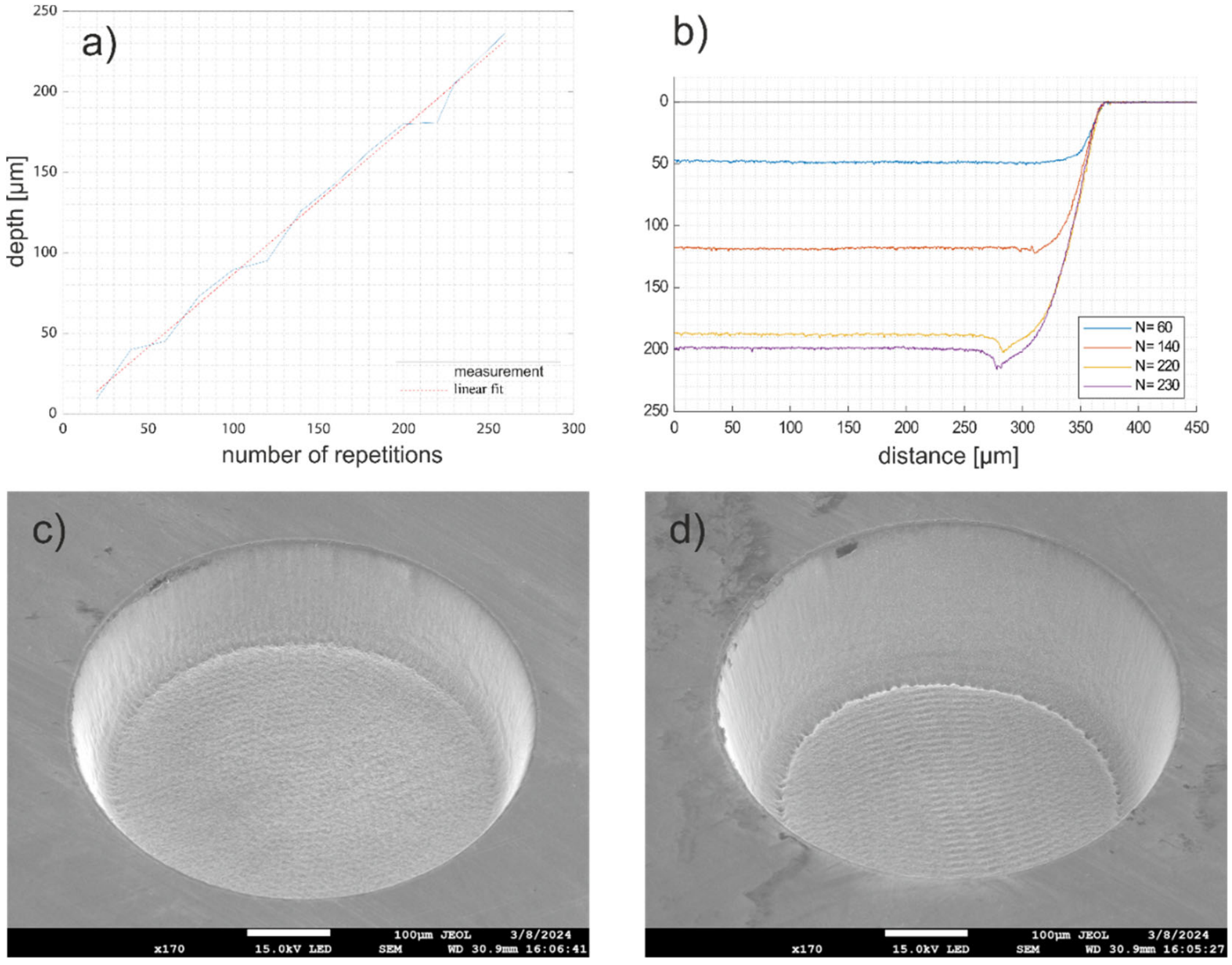

**Fig. 2** Panel **(a)** shows the linear dependence of the depth removed per scan repetition. Measurements of the cavity profile for different repetition values using a confocal microscope show indentations at the transition from the base to the side edge **(b)**. This becomes *even* more pronounced as the number of repetitions increases. Panels **(c)** and **(d)** show SEM images of the cavity at N=140 and 220, respectively. After 140 repetitions, the first signs of irregularities forming along the side edge can already be seen. After 220, a clear ring with significant indentations has formed.

In order to investigate the profile geometry, measurements using a Laser Scanning Microscope (LSM) have been carried out, the results can be seen in panel b of Fig. 2. The cavities were measured after machining with a selection of passes at N = 60, 140, 220 and 230. In all cases one can see that the bottom plane is parallel to the substrate surface, indicating uniform removal. The transition from the ground to the side edge forms a radius due to the Gaussian beam profile, as clearly visible after a repetition number of N=60. An increased roughness in this area can already be recognized in the measurement at N=140. After N=220 and higher, one can clearly see pits in this transition region. Panels c and d of Fig. 2 show SEM images of the cavities created with N = 140 and N = 220, respectively. An increased roughness in the measurement and the beginning of the formation of small holes along the sideline are visible. By increasing the number of repetitions, these holes become even more pronounced and form a recessed ring around the bottom surface. This effect is well known and has already been reported [27], [28] . It is most likely caused by reflections at the side edge, as discussed by Toenshoff and co-authors during their research on femtosecond silicon cutting [29]. The reflections lead to an increased intensity in the edge area and the formation of the indentation. The profile measurements also show that the side walls form an inclination angle of approximately 13°, despite using the same diameter for each repetition. This can be attributed to the increase of the laser beam's area projected on the side walls at larger depths, which in turn reduces the fluence to the level of the threshold fluence, hence no further ablation occurs [30].

To mitigate the sideline hole formation described above and to systematically examine how different scanning patterns affect the resulting cavity geometry, the following processing strategies were employed:

- large to small (LTS), where the diameter of the circular hatch pattern is decreased for each repetition
- small to large (STL), where the diameter is increased for each repetition
- random (RND), where the diameter is randomly chosen between a maximum and minimum value for each repetition

Each strategy was processed with the same parameters listed in table 1.

|  |  |
|---|---|
| Wavelength | 523 nm |
| Lens, focal length | 170 mm |
| Repetition rate | 100 kHz |
| Markspeed | 1 m/s |
| Power | 0.45 W @ 100 kHz |
| Focusshift | -0.75 μm for each repetition |
| Polarisation | Circular |
| Cavity diameter | 500 μm |

**Table 1** Process parameters for realising the thinned areas. The values have already been optimized based on the results of previous publications. The marking speed and repetition rate result in a pulse-to-pulse distance of 10 microns. This corresponds approximately to the beam radius and yields the lowest surface roughness during ablation. Furthermore, the power value lies within the range of maximum ablation efficiency and is therefore the ideal value for gentle thinning with highest surface quality.

All thinning strategies (LTS, STL and RND) were followed with the same parameters from table 1 using 230 repetitions, which led to cavities with a depth of 200 μm. Just the diameter of the circle was changed for each repetition. Thus, the difference in surface quality and geometry is solely attributed to the scanning strategy applied.

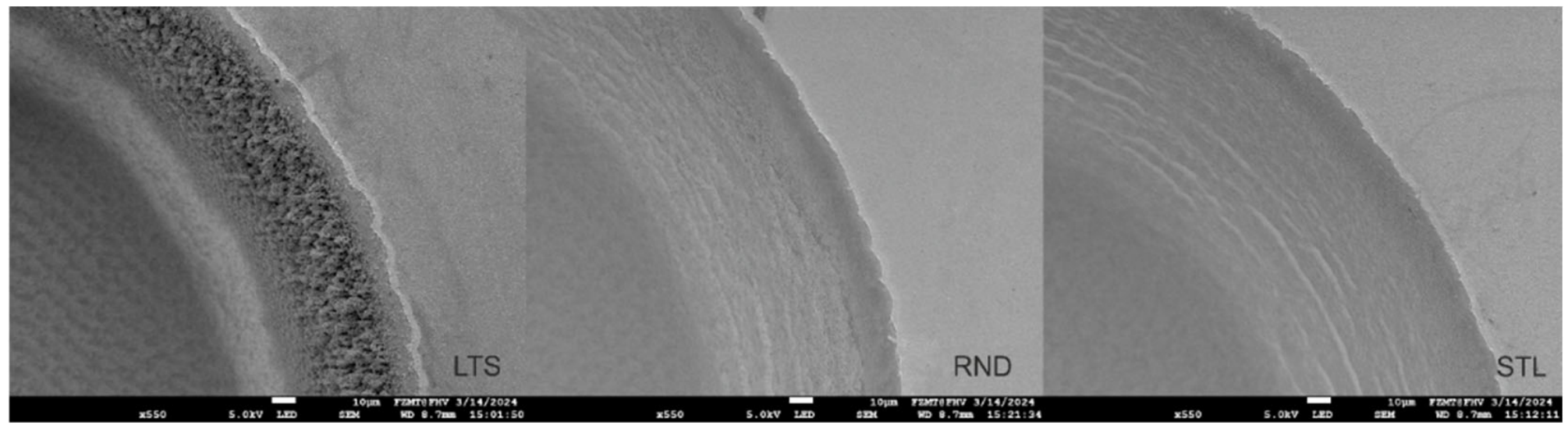

**Fig. 3** SEM images of sections of cavities processed with three different scanning strategies. On the left image (LTS) the diameter of the ablation pattern decreases with every pattern repetition. In case of RND the diameter of the pattern is randomly chosen between two values and in case of STL the diameter increases during processing.

The inner and outer diameters (450 µm and 500 µm) were chosen in a way that they correspond to the inclination angle of the side wall that could already be observed in the previous tests. Fig. 3 shows SEM images of sections of the cavities. The transition from the bottom surface to the side wall shows that the formation of the holes and trenches was suppressed with all three methods. This confirms the assumption that the cause of their formation are reflections of the laser beam at the side edge. In all three variants, the passing over this area was reduced. In the area of the side edge, a deterioration in the edge quality can be clearly recognised in the case of LTS. The edge shows a high surface quality in the lower area of the cavity. Moving upwards, an increasing amount of deposits accumulate towards the upper edge. By reducing the scanning area after each repetition, the debris are deposited more and more in the outer areas. With the RND and STL methods, these deposits cannot be observed as any existing debris is constantly removed from the side edge by repetitions with a larger diameter. In general, the RND and STL methods show similar morphologies with comparable surface roughness.

With these two selected strategies, RND and STL, experiments were conducted where the inner diameter was varied for each cavity. These span from 350 µm to 450 µm with steps of 20 µm, with the aim to examine possible variations of the side wall angle. This is particularly important for applications of PMN-PT as actuators where electrical contacts are realised through vapor deposition of a thin metal layer on the surface. Here, the side wall angle and the surface quality have a decisive impact on the conductance of the layer in these critical areas.

The cavity profiles realized in this way were again measured using the laser scanning microscope. The results can be seen in the graphs of Fig. 4. In general, only a slight change in the side wall angle can be observed as a function of the inner diameter. The values deviate significantly from the ones expected from geometric calculations and are comparable for both methods. A possible explanation for the low variation in flank angles is that for low changes in diameter (less than 1 micron) the angles and the profile are influenced predominantly by the gaussian profile of the beam rather than geometrical parameters. As discussed above, the side-wall angle originates from the progressive increase of the laser beam's projected footprint on the trench side walls with increasing ablation depth. The enlarged incidence area leads to a corresponding reduction in local fluence, which approaches the material's threshold fluence and ultimately terminates the ablation process [31], [32].

A difference between the RND and the STL method can be observed in the formation of the bottom surface.

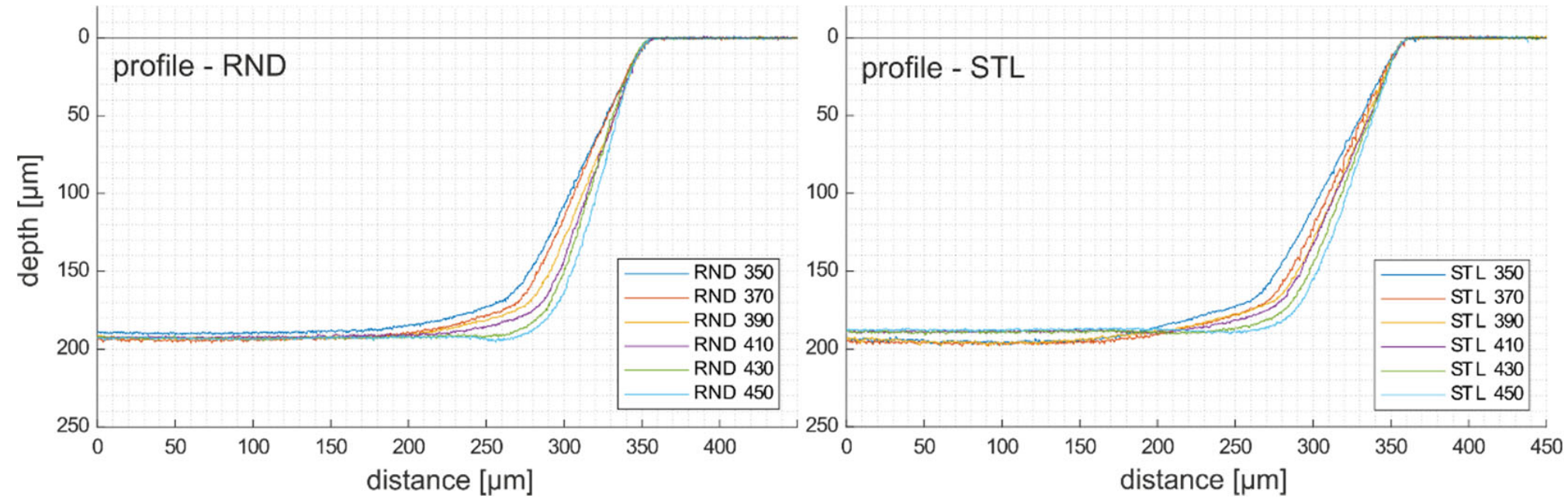

**Fig. 4** Cavity profiles obtained with the laser scanning microscope for the RND and STL methods. Both techniques show only minor variations in side-wall angle with changing inner diameter, and the measured values deviate notably from geometrical expectations.

For parameters that lead to steeper edge angles, RND shows a slight indentation in the corner region, whereas with the STL method, the entire bottom surface shows a uniform depth. However, with smaller inner diameters, this method shows a significantly more curved bottom surface compared to RND. To further examine the possible variation in flank angles towards significantly lower values, cavities were generated by applying the same parameters as in table 1 but with the chosen geometry altered. Instead of using a circle, a rectangle with 500 microns in width and 2.5 mm in length was chosen. The rectangle was then shifted by a fixed value from left to right for each of the 200 repetitions. The values range from 0 to 10 µm, with a step of 1 micron. In total, 11 cavities were generated with this method. By using this scheme, the left side resembled LTS, while the right side resembled STL. Fig. 5 shows the results for the left and right side of 3 cavities with a shift value of 1, 2 and 5 microns, respectively.

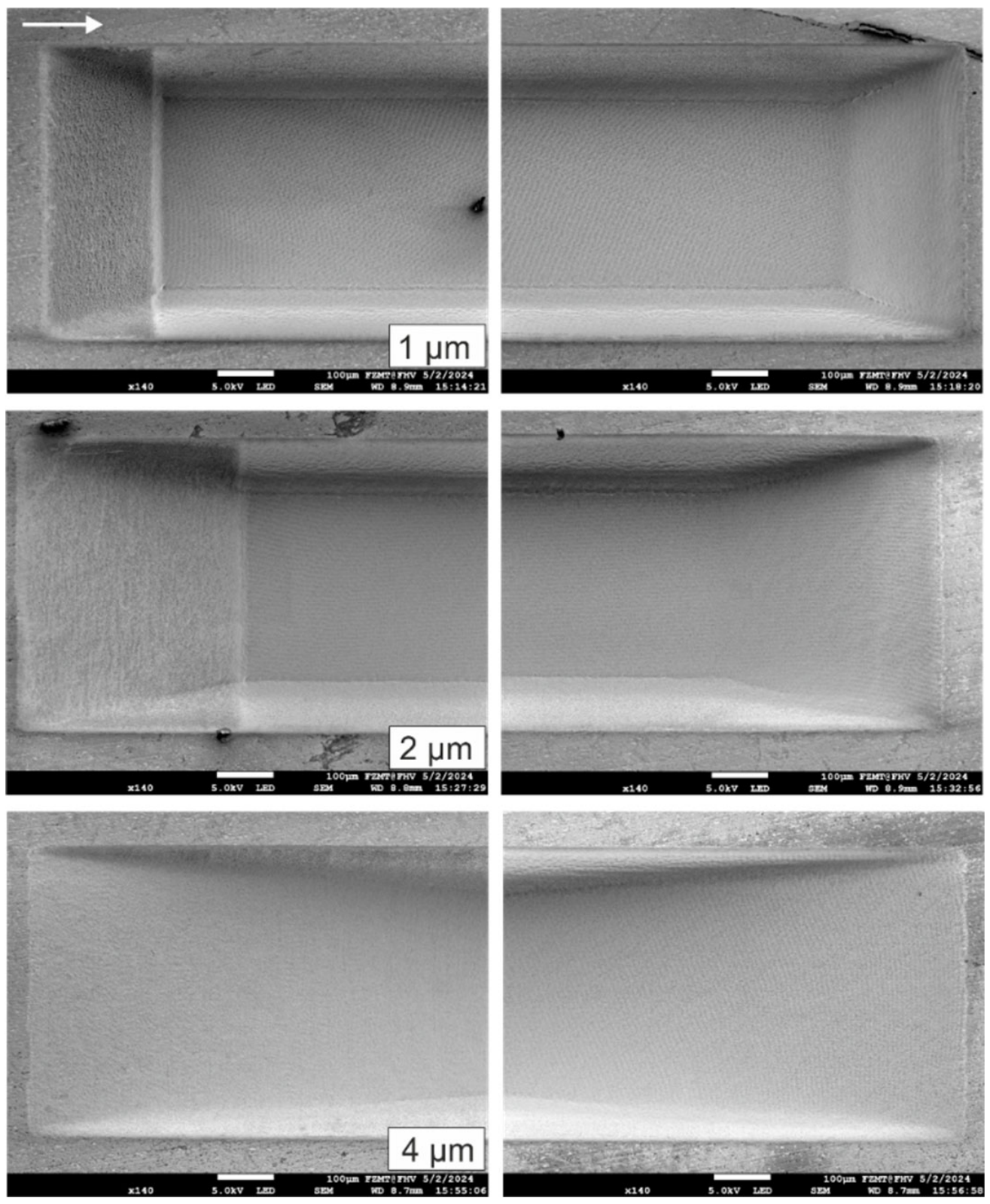

**Fig. 5** SEM images of left and right ends of cavities realised with shift values of 1 µm (top), 2 µm (middle) and 4 µm (bottom line). *I*t is clearly visible that the flank angles can be varied in a larger range than the previous experiments show.

As moving the pattern to the right means that the machined surface on the left is no longer passed by the laser, debris is expected to be deposited on the side edge. The reduced surface quality can be recognised at a pattern shift value of 1 µm. However, the amount of accumulations is already reduced significantly compared to the previous tests, which correspond to a shift of 0.11 µm. The right ends show a good quality for all parameters, similar to STL. Additionally, it is clearly visible that the flank angles can be varied in a larger range than the previous experiments show. The corresponding profile measurements are presented in the left panel of Fig. 6.

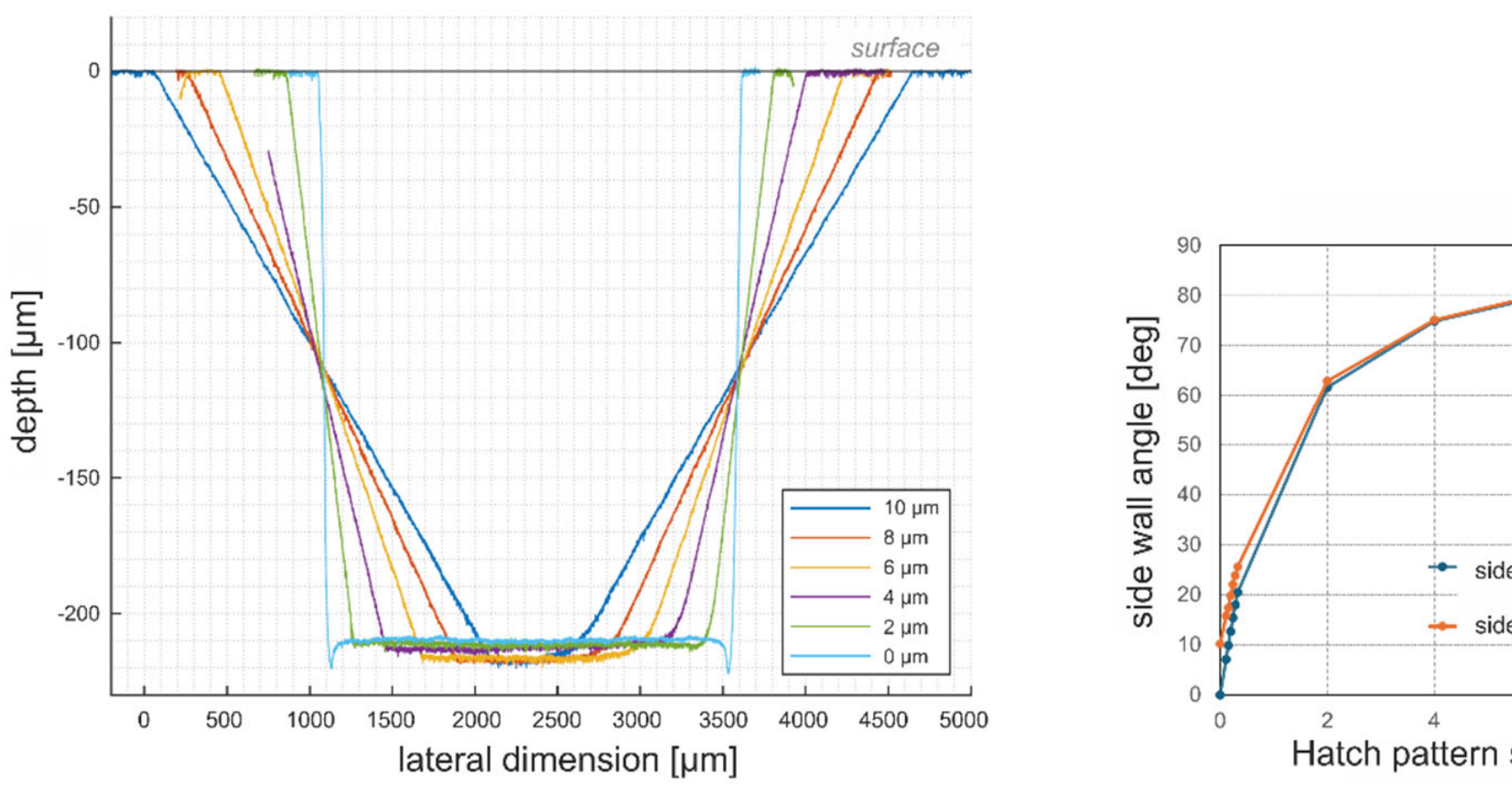

**Fig. 6** left: Measurement of the cavity profile for different values of pattern shift showing straight lines with no influence of the shift direction on the side wall angle; right: A comparison of the measured side wall angle with the estimated value. The angles correspond for offset values higher than 4 µm.

The profile measurements demonstrate clearly that it is possible to generate cavities with a larger variety in flank angles. It can also be seen that, if the pattern is moved constantly, an inclined plane of the same angle is formed on both the left and right side. In addition, the resulting angle corresponds exactly to the calculated value for pattern shift values higher than 4 µm (Fig. 6 right graph). Above this value, the influence of the Gaussian beam profile which leads to higher angles as expected is no longer observable. This makes it possible to realize quasi-3D structures (limited to non-undercut geometries) by laser thinning with adapted hatch layers. Since the geometry of the side wall in terms of shape and angle is independent of the processing method, there is a clear difference in case of the bottom edge. A rounded edge forms on the left which is of comparable dimensions as the laser beam. On the right, there is a smooth transition from the bottom surface to the side wall with a significantly higher radius (Fig. 6 left graph). We assume that the reason for the difference in radius lies in the different geometrical starting points. When the rectangle is shifted to the right with each repetition, the ablation on the left side always occurs on a flat surface, finally creating a sharp edge. Contrary to that, the ablation on the right side always occurs over the end of the previous ablated surface and thus on an uneven surface condition. This leads to ablation with reduced fluence in this area, which does not affect the final side wall angle but does lead to a rounding of the geometry at the bottom edge.

## 3.2 Realisation of conducting paths beyond laser thinned areas

To electrically control all, even thinned areas of the piezo, we must be able to apply a continuous conductive layer over the previously processed surfaces. Since the non-even metallization layer cannot be patterned easily by standard lithography methods, it is subsequently structured by means of selective laser ablation to form individual areas for actuator control. To prepare the substrate for evaporation tests, it was first thinned using the parameters for the circular hatch pattern with a variation of the inner diameter. After a cleaning step with deionized water and subsequent treatment in oxygen plasma the samples were coated with titanium in steps. During each coating process, 25 nm with a deposition rate of 1.5 Å $s^{-1}$ were deposited onto the surface while the substrate was rotated within the machine. For final applications, a gold coating is preferred due to its significantly higher conductivity. However, these tests were carried out with Ti due to its higher adhesion. This reduces the risk of delamination of the individual layers as the process was interrupted several times for the conductivity measurement. The resistance was measured using tungsten needles, which are connected to a multimeter; one was placed on the bottom of the cavity, while the other one was placed at the top surface. The resistance was measured between these two points, the results for both thinning strategies can be seen in Fig. 7.

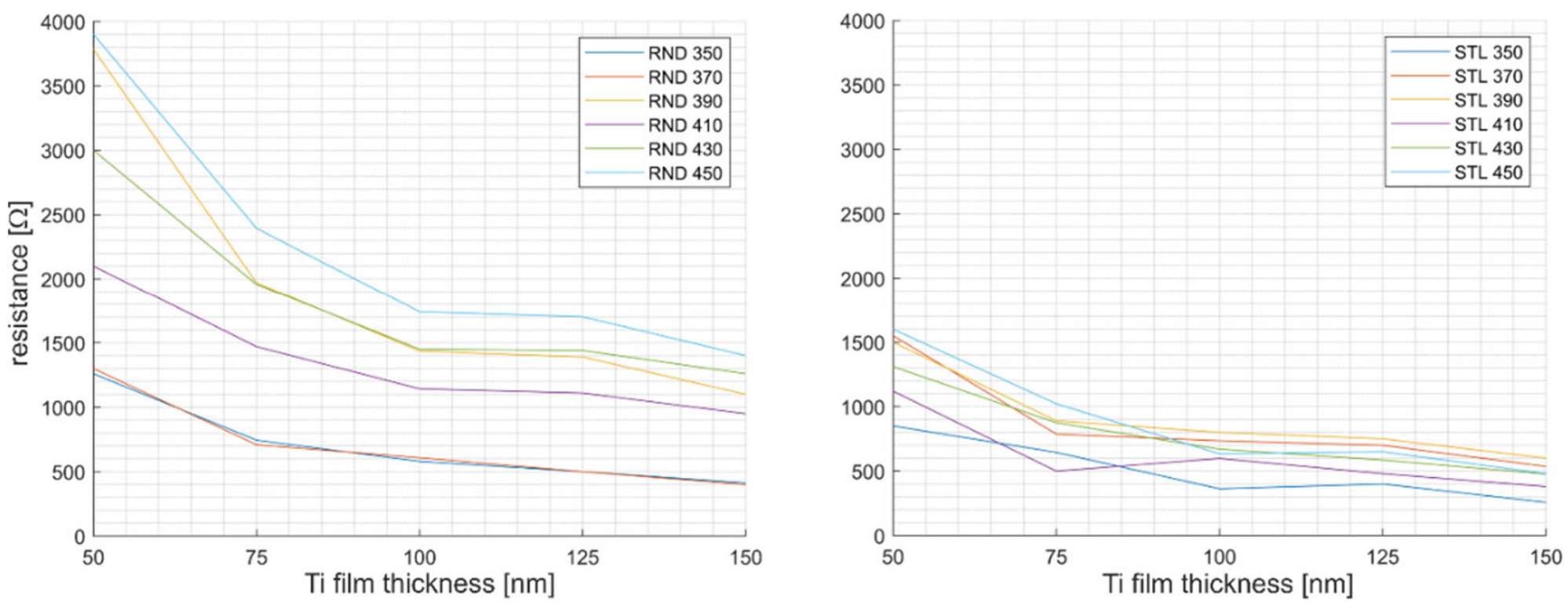

**Fig. 7** Electrical resistance measurements for various layer thicknesses of Ti for cavities generated with varying inner diameters (indicated in the legend) using the RND (image on the left) and STL (image on the right) method.

Although the methods RND and STL showed comparable results regarding morphology (Fig. 3), the slight differences in edge geometry leads to significant variations in the conductance of the deposited metallic layer. Particularly in the case of steeper edges, the cavities realized with the RND method exhibit considerably higher values of electrical resistance (Fig. 7 left). While these values decrease as the layer thickness increases, they remain higher compared to STL (Fig. 7 right). The difference is likely due to the noticeable indentation at the side edges in the case of RND, which seems to have a direct impact on the layer conductivity in this region. As a result of these measurements further examinations regarding layer structuring were continued with cavities fabricated using the STL method.

Circular cavities were generated for the tests to create structured metallic contacts on non-planar surfaces. Inner diameters of 350 µm and 450 µm were chosen, resulting in the smallest and largest values for the side wall inclination, as demonstrated in the right graph of Fig. 4. After a cleaning step, a coating of 10 nm titanium and 150 nm gold was applied. After that, a rectangular structuring path was ablated using the femtosecond laser, with one of the long sides placed centrally over the cavity (see schematic of Fig. 8). After structuring, the electrical separation between the inner and outer surfaces of the rectangle was tested both on the cavity floor and on the substrate surface. In addition, the conductivity from the bottom to the top surface was also measured as in the coating tests.

One characteristic of ablation with ultrashort laser pulses is a sharply defined ablation threshold. This was measured using Liu's method [26] and is 0.14 J/cm² in the case of gold and 1 J/cm² for PMN-PT at a wavelength of 523 nm. These values show the potential for selective structuring of gold on PMN-PT, which we demonstrate below. For the layer structuring, a repetition rate of 100 kHz and a processing speed of 700 mm/s were selected. This results in a slightly higher pulse overlap, which suppresses the visible pattern of individual ablation spots and produces a cleaner, straighter edge of the gold layer.Different samples were structured with power values of 0.882 W and 0.788 W, all with 2 and 3 pattern repetitions. The focus was positioned to half the depth of the cavity. This provides the highest possible value for the intensity at the side edge. This is advantageous as the fluence value is smaller compared to flat surfaces due to projection onto an inclined plane. The Rayleigh length of the focused beam is 696 µm and is considerably higher than the depth of the structure.

Electrical resistance between the separated regions, both on the bottom of the trench and on the substrate surface, must be very high. Measurements show that all eight combinations of laser power and repetition number yield resistance values exceeding the multimeter's upper measurement limit of 50 MΩ. The resistance of the gold layer across the structured side edge, as examined in the coating tests, was also measured and found to be between 3 Ω and 9 Ω for all samples. The SEM image in Fig. 8 shows the result of the structuring process at a laser power of 0.882 W and three repetitions. The slight contrast difference inside and outside the rectangle, which is due to different electrical charges, already indicates the distinct electrical properties of these regions. The measurement positions for conductivity are shown schematically (Fig. 8 right).

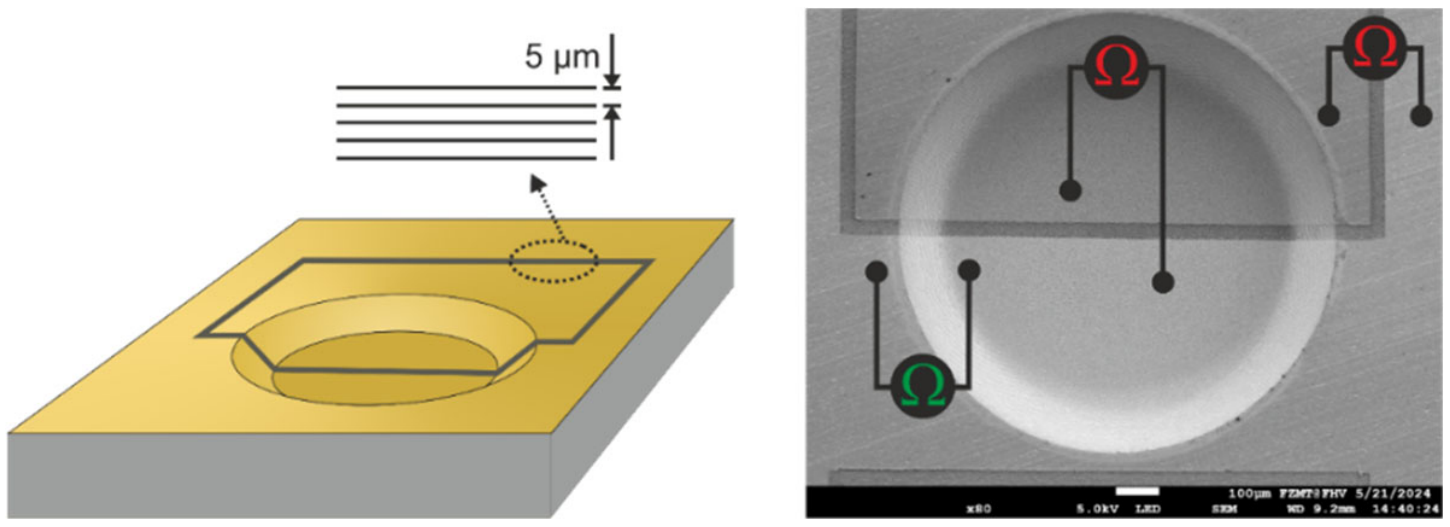

**Fig. 8** Schematic of the laser-structured Au-layer (left), showing the rectangular separation track placed beyond the cavity. To obtain a certain path width, 5 parallel lines with 5 µm distance were applied. The laser power was 0,88 W and the pattern was repeated three times. The SEM image on the right confirms complete removal of the thin gold layer. The slight contrast difference inside and outside the rectangle, caused by different electrical charges, indicates the electrical isolation of the two regions. The measurement points for conductivity determination are also marked in the SEM image.

## 3.4 UV laser cutting within the locally thinned areas

PMN-PT based piezo platforms realized by laser cutting for quantum optical applications has already been successfully demonstrated [33], [34], [35]. The cutting process was based on the creation of equidistant lines to the cutting path geometry with variable line spacings. Process parameters like fluence and repetition rate were within a narrow process window allowing cutting with minimal thermal stress on this very sensitive material. The laser was operated at a wavelength of 523 nm to optimize absorption and reduce collateral damage. However, when moving toward the significantly miniaturized actuator designs pursued in this work, these established process parameters are no longer sufficient to meet the required precision. Consequently, an adapted processing strategy is introduced here.

As a first change in the parameter set, the wavelength of the laser was reduced to the third harmonic at 348 nm. This provides the advantage of a smaller spot radius on the one hand and improved absorption conditions on the other. Data from the literature determine the bandgap of monocrystalline PMN-PT with a value in the range of 3.1 to 3.3 eV, depending on material composition and temperature [36]. This is just between the photon energy of the laser at its second (2.37 eV) and third harmonic (3.56 eV), respectively. Using the UV wavelength, the laser energy can be deposited via linear absorption, in contrast to the previously employed cutting process, which relied on two-photon absorption. This offers the advantage of intensity-independent energy deposition within the optical penetration depth and, consequently, reduces the impact on the surrounding bulk material. As already mentioned, the refined actuator concepts are based on a locally thinned area in which a miniaturised design is realised using the UV fs-laser. However, as PMN-PT is difficult to manipulate mechanically at this greatly reduced thickness of 100 µm, a substrate with a thickness of 300 µm was used for the first tests to evaluate the cutting edge. To refine the process parameters and evaluate the laser exit side, the substrates were thinned to the reduced thickness of 100 µm using the process described in section 3.1.

The SEM images in Fig. 9 show the design of a three-leg actuator, realised with different power values and variable number of passes. Two sets of equidistant lines were applied alternately to achieve the required cutting depth. First, 10 lines were applied with a distance of 4.5 µm to ensure removal even at the wider parts of the geometry, so that there were no free-standing parts at the end of the cutting process. Then, additional equidistant lines with reduced distance (5 x 2 µm) should lead to a more distinctive cutting edge.

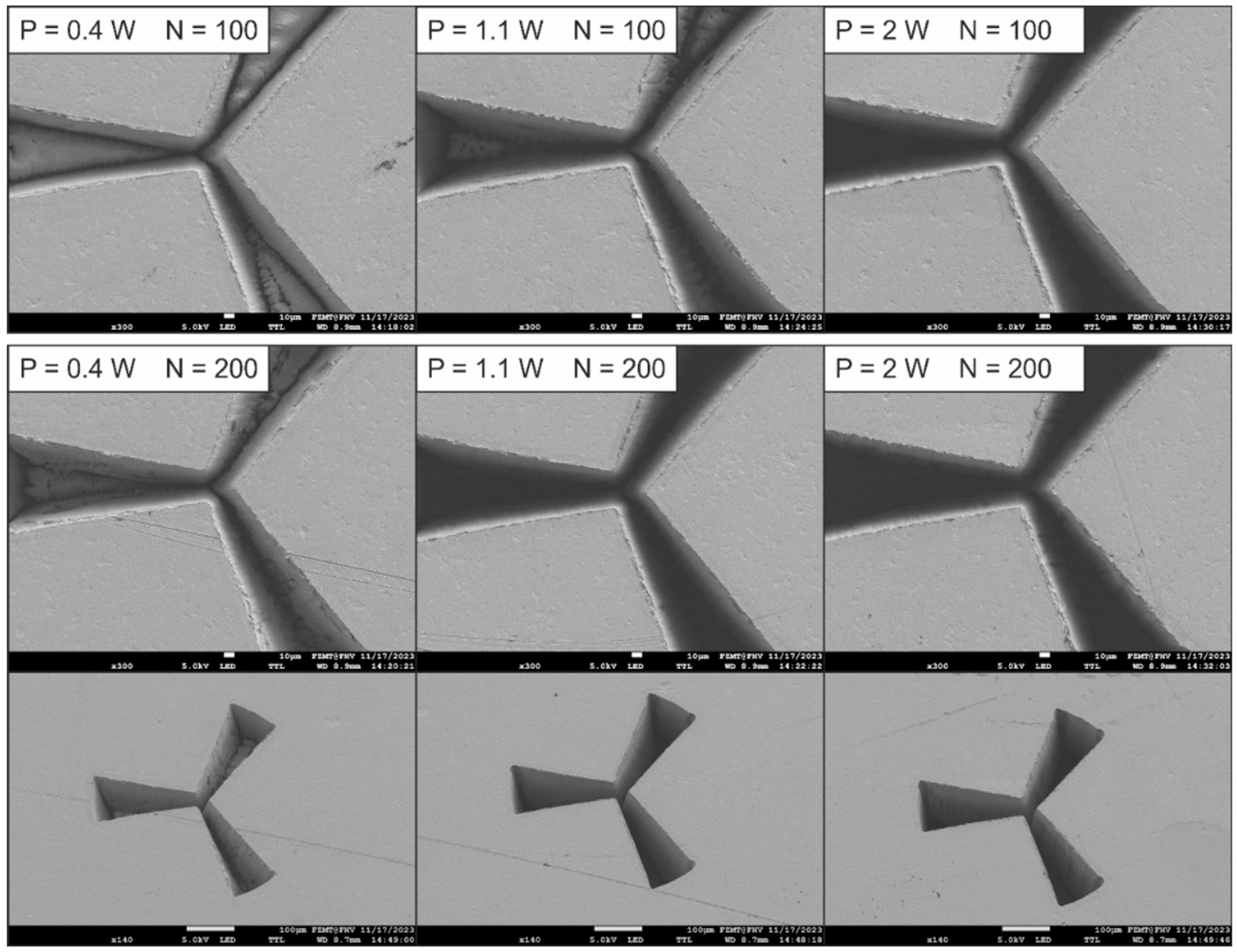

**Fig. 9** SEM images of a three-leg actuator design cut with different laser and process parameters using the third harmonic wavelength. For the process, two sets of equidistant lines were applied; 5 lines with a distance of 2 µm and 10 lines with a distance of 4.5 µm. In the upper row results for three different values of laser power and a repetition number of 100 is shown. Row 2 and 3 show results after 200 repetition and an additional cutting design overview.

The laser was operated at a repetition rate of 250 kHz. This was chosen to allow the laser to operate at its maximum pulse energy, as the IR-to-UV frequency conversion stage is specifically optimized for this setting. The scanner speed was 1250 mm/s for all tests. The results shown in the upper row were obtained with a repetition of the cutting line pattern of 100. It can be clearly seen that the ablation depth is very low, particularly at a power of 0.4 W. Although this power represents the optimal operating range with the highest ablation efficiency (see Appendix A.2, ablation efficiency measurements), the resulting depth at 0.4 W is insufficient for performing a complete cut. Therefore, the laser power must be increased beyond the efficiency optimum to achieve full through-cutting, even though this requires operating outside the ideal parameter range. Raman measurements performed in the cavities used for determining the ablation efficiency showed no detectable influence of the processing parameters on the Raman spectra even at these higher power levels. Further details are provided in the Appendix A.3.

The arrangement of the cutting lines produces a cushion-shaped base within the cutting geometry, resulting in a pronounced depression along the intended cutting path. With increasing laser power, the ablation depth rises significantly, while the edge quality shows almost no change. At a repetition number of 200, the cushion-like base is only visible at the lowest power of 0.4 W. For higher values, the edge progression can no longer be reliably evaluated. Examination of the cutting edge on the laser-entrance side reveals high quality without any indications of chipping or breakouts at 1.1 W. Since slight melting becomes apparent at 2 W, subsequent experiments are carried out at a power of 1.1 W.

To estimate the quality to be expected on the exit side, the bottom of the cavity for a power of 1.1W at 100 repetitions can be evaluated. This already shows increasing irregularities with depth, which suggests that an adaptation of the machining strategy is required to improve the quality. In this alternative approach, the actuator geometry is not cut by generating equidistant lines but instead processed using a line-hatch pattern with the parameters described in Section 3.1. This enables pre-thinning of the geometry with the highest surface quality and a flat inner area. SEM images in Fig. 10 show the actuator geometry during thinning for different values of pattern repetition. Due to the small dimensions and the Gaussian beam profile, a rounded cavity shape forms. Starting from a repetition of 15, slight waviness appears at the bottom of the structure. Above a repetition of 70, an irregular side edge forms at the bottom of the structure. As a lower fluence is used when thinning compared to cutting, a cavity with a lower side edge angle is created. As already discussed in [18], the angle of the side edge depends solely on the ratio of the laser fluence to the ablation threshold of the material. It is expected that the higher fluence during cutting will widen the cavity, which should have a positive effect on the cut edge definition. After a defined number of repetitions, the cutting procedure switches to the actual cutting process with a set of equidistant lines and higher fluence values. This is intended to improve the quality of the final cutting edge on the laser entrance as well as on the laser exit side.

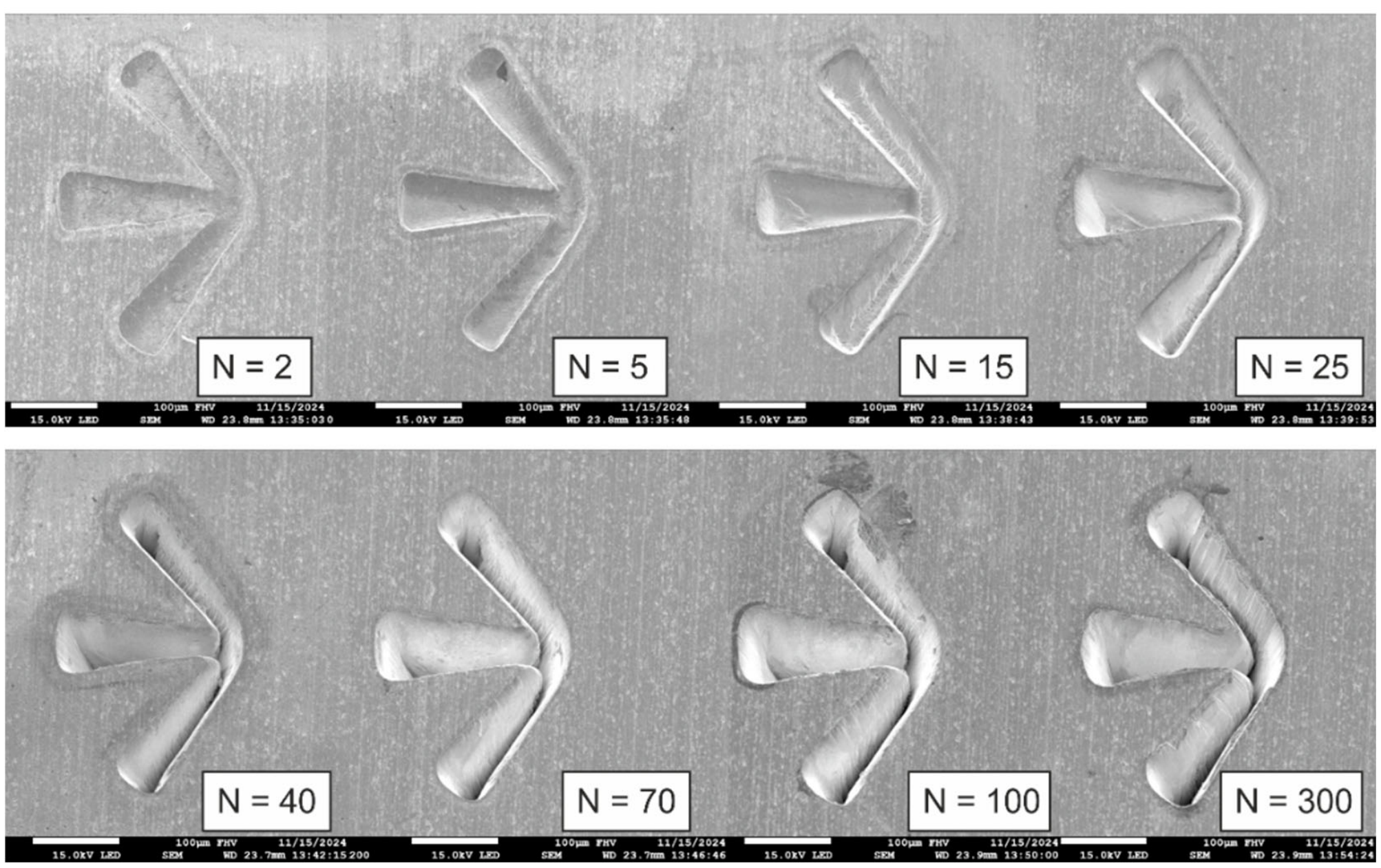

**Fig. 10** SEM images of cavities generated by a line hatch pattern for an increasing number of pattern repetitions. As the number of repetitions increases, a rounded mould first forms, which deepens more and more and develops into a cut edge for repetitions higher than 40. A trough formation in the ablation is already recognisable from the 5th repetition, which transforms into a trench from the 15th repetition. The lateral flanks become increasingly steep and misshapen.

In preparation for the final cutting tests, circular cavities with a residual thickness of 100 µm were fabricated. Within these cavities, a geometrically divided actuator, each half equipped with two independently controllable arms, is realised. For the processing steps, the substrate is bonded to a ceramic sample holder using polyvinyl alcohol (PVA). Particular care is taken to ensure that both the front and the back side of the piezo substrate are fully covered with the coating. This is necessary to enable the removal of debris from the cutting process together with the PVA layers.

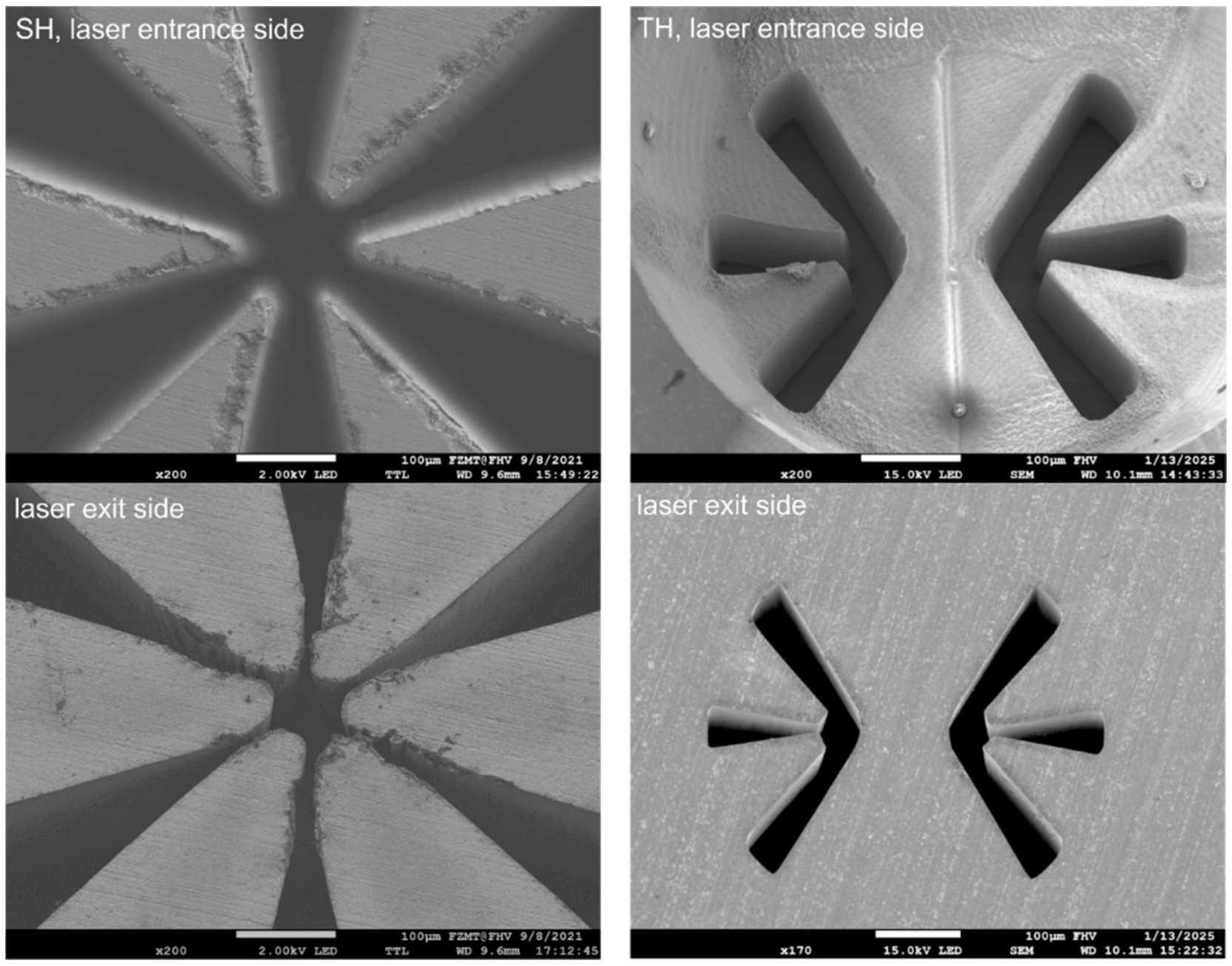

**Fig. 11** SEM images of laser-cut actuators processed with the second harmonic (left) and the third harmonic wavelength (right) clearly demonstrate the quality improvement achieved by using UV radiation. Both devices were fabricated with optimized process parameters; in the UV case, the actuator was structured within a laser-thinned region with a residual substrate thickness of 100 µm. The minimum inter-arm spacing on the laser-exit side is only 5 µm, representing a substantial reduction compared to designs produced with the SH wavelength and directly contributing to higher achievable strain values. In addition, the absence of irregularities on the back side of the UV-processed actuator is highly advantageous for subsequent fabrication steps such as bonding.

Fig. 11 presents the front and back sides of two actuators. The actuator shown in the left column was cut using the second harmonic wavelength; the corresponding parameter study on optimized cutting conditions is reported in [18]. In contrast, the actuator in the right column was fabricated using the UV wavelength with the above-mentioned cutting procedure where the preliminary thinning of the actuator geometry (N = 100) was followed by the final cutting step with equidistant lines (N = 200). The cut was positioned within a circularly thinned region with a residual substrate thickness of 100 µm. Although deposition residues are still visible inside the cavity - most likely caused by insufficient PVA coverage - the overall edge quality is markedly improved. Compared to the actuator on the left, the UV-processed device exhibits sharp and well-defined edges on both sides. The exit side is particularly critical for subsequent processing steps, as this is where the semiconductor membrane containing the quantum emitters will be integrated. The SEM image in the lower right confirms that the UV-based process produces clean edges without protrusions or breakouts, representing a substantial improvement over the second-harmonic results. Furthermore, the minimum cut width on the exit side is reduced to approximately 5 µm, compared to the 10–15 µm typically observed with second harmonic (SH) processing. This significant narrowing is expected to enable higher strain values at lower operating voltages than those achievable with previously used tuning platforms.

## 5. Conclusion

In this work, we successfully demonstrated the fabrication of miniaturized and structurally complex piezo actuators based on single crystal PMN-PT using exclusively ultrashort pulsed laser processing for local sample thinning, electrode structuring and actuator cutting. In particular, the use of UV wavelengths enabled highly precise laser cutting, which proved essential for achieving the required resolution and structural precision. The newly developed process sequence enables actuator designs that offer promising new approaches for applications in the field of multiaxial strain-tuning, such as the integration of multiple individually addressed quantum emitters or the realisation of higher and more directional strain values with miniaturized devices.

## Acknowledgements

Parts of the work was funded within the QuantERA Programme that has received funding from the EU H2020 research and innovation programme under GA No 101017733 via the project QD-E-QKD and from the MUR (Ministero dell'Università e della Ricerca) through the PNRR MUR Project PE0000023-NQSTI and via the Project MEEDGARD: grant.13948955, EPSRC EP/Z000556/1 and 425333704 FWF-DFG.

J.M.-S. acknowledges financial support from the Spanish Ministry of Science and Innovation (grant number PID2023-148457NB-I00 funded by MCIN/AEI/10.13039/501100011033 and FSE+, PCI2022-132953 funded by MCIN/AEI/10.13039/501100011033 and the EU "NextGenerationEU"/PRTR, CNS2024-154342 funded by MICIU/AEI /10.13039/501100011033")

# References

[1] J. Holterman und P. Groen, *An introduction to piezoelectric materials and applications*. Apeldoorn: Stichting Applied Piezo, 2013.

[2] S.-E. Park und T. R. Shrout, „Ultrahigh strain and piezoelectric behavior in relaxor based ferroelectric single crystals", *J. Appl. Phys.*, Bd. 82, Nr. 4, S. 1804–1811, Aug. 1997, doi: 10.1063/1.365983.

[3] S.-H. Baek, M. S. Rzchowski, und V. A. Aksyuk, „Giant piezoelectricity in PMN-PT thin films: Beyond PZT", *MRS Bull.*, Bd. 37, Nr. 11, S. 1022–1029, Nov. 2012, doi: 10.1557/mrs.2012.266.

[4] E. Sun und W. Cao, „Relaxor-based ferroelectric single crystals: Growth, domain engineering, characterization and applications", *Prog. Mater. Sci.*, Bd. 65, S. 124–210, Aug. 2014, doi: 10.1016/j.pmatsci.2014.03.006.

[5] T.-B. Xu, „Review on PMN-PT Relaxor Piezoelectric Single Crystal materials for cryogenic actuators", in *AIAA SCITECH 2022 Forum*, San Diego, CA & Virtual: American Institute of Aeronautics and Astronautics, Jan. 2022. doi: 10.2514/6.2022-2240.

[6] J. Peng, C. Chao, J. Dai, H. L. W. Chan, und H. Luo, „Micro-patterning of 0.70Pb(Mg1/3Nb2/3)O3–0.30PbTiO3 single crystals by ultrasonic wet chemical etching", *Mater. Lett.*, Bd. 62, Nr. 17–18, S. 3127–3130, Juni 2008, doi: 10.1016/j.matlet.2008.02.003.

[7] I. A. Ivan, J. Agnus, und P. Lambert, „PMN–PT (lead magnesium niobate–lead titanate) piezoelectric material micromachining by excimer laser ablation and dry etching (DRIE)", *Sens. Actuators Phys.*, Bd. 177, S. 37–47, Apr. 2012, doi: 10.1016/j.sna.2011.09.015.

[8] C. Liu, F. Djuth, X. Li, R. Chen, Q. Zhou, und K. Kirk Shung, „Micromachined high frequency PMN-PT/epoxy 1–3 composite ultrasonic annular array", *Ultrasonics*, Bd. 52, Nr. 4, S. 497–502, Apr. 2012, doi: 10.1016/j.ultras.2011.11.001.

[9] J. Zhang, W. Ren, X. Jing, P. Shi, und X. Wu, „Deep reactive ion etching of PZT ceramics and PMN-PT single crystals for high frequency ultrasound transducers", *Ceram. Int.*, Bd. 41, S. S656–S661, Juli 2015, doi: 10.1016/j.ceramint.2015.03.258.

[10] Y. Chen *u. a.*, „Addressable and Color-Tunable Piezophotonic Light-Emitting Stripes", *Adv. Mater.*, Bd. 29, Nr. 19, S. 1605165, Mai 2017, doi: 10.1002/adma.201605165.

[11] Z. Lei *u. a.*, „Micromachining of High Quality PMN–31%PT Single Crystals for High-Frequency (>20 MHz) Ultrasonic Array Transducer Applications", *Micromachines*, Bd. 11, Nr. 5, S. 512, Mai 2020, doi: 10.3390/mi11050512.

[12] J. Lv *u. a.*, „Cold ablated high frequency PMN-PT/Epoxy 1-3 composite transducer", *Appl. Acoust.*, Bd. 188, S. 108540, Jan. 2022, doi: 10.1016/j.apacoust.2021.108540.

[13] K. H. Lam, Y. Chen, K. Au, J. Chen, J. Y. Dai, und H. S. Luo, „Kerf profile and piezoresponse study of the laser micro-machined PMN-PT single crystal using 355nm Nd:YAG", *Mater. Res. Bull.*, Bd. 48, Nr. 9, S. 3420–3423, Sep. 2013, doi: 10.1016/j.materresbull.2013.05.025.

[14] K. Sugioka, M. Meunier, und A. Piqué, Hrsg., *Laser precision microfabrication*. in Springer series in materials science, no. 135. Berlin Heidelberg Dordrecht: Springer, 2010.
[15] G. Dumitru, V. Romano, H. P. Weber, M. Sentis, und W. Marine, „Femtosecond ablation of ultrahard materials“, *Appl. Phys. Mater. Sci. Process.*, Bd. 74, Nr. 6, S. 729–739, Juni 2002, doi: 10.1007/s003390101183.
[16] M. Farsari, G. Filippidis, S. Zoppel, G. A. Reider, und C. Fotakis, „Efficient femtosecond laser micromachining of bulk 3C-SiC“, *J. Micromechanics Microengineering*, Bd. 15, Nr. 9, S. 1786–1789, Sep. 2005, doi: 10.1088/0960-1317/15/9/022.
[17] J. Song *u. a.*, „Piezoelectric micropatterns fabricated by femtosecond laser processing for *d* 33-mode piezoelectric MEMS“, *J. Appl. Phys.*, Bd. 139, Nr. 4, S. 044102, Jan. 2026, doi: 10.1063/5.0304045.
[18] G. Piredda *u. a.*, „Micro-machining of PMN-PT crystals with ultrashort laser pulses“, Bd. 125. Jg., Nr. H. 3, S. 11, 2019, doi: 10.1007/s00339-019-2460-9.
[19] J. Martín-Sánchez *u. a.*, „Reversible Control of In-Plane Elastic Stress Tensor in Nanomembranes“, *Adv. Opt. Mater.*, Bd. 4, Nr. 5, S. 682–687, Mai 2016, doi: 10.1002/adom.201500779.
[20] R. Trotta *u. a.*, „Wavelength-tunable sources of entangled photons interfaced with atomic vapours“, Bd. o.Jg., Nr. Nr. 10375, S. 7, 2016, doi: 10.1038/ncomms10375.
[21] J. Yang, M. Zopf, und F. Ding, „Strain tunable quantum dot based non-classical photon sources“, *J. Semicond.*, Bd. 41, Nr. 1, S. 011901, Jan. 2020, doi: 10.1088/1674-4926/41/1/011901.
[22] D. Huber *u. a.*, „Strain-Tunable GaAs Quantum Dot: A Nearly Dephasing-Free Source of Entangled Photon Pairs on Demand“, *Phys. Rev. Lett.*, Bd. 121, Nr. 3, Juli 2018, doi: 10.1103/physrevlett.121.033902.
[23] J. Martín-Sánchez *u. a.*, „Strain-tuning of the optical properties of semiconductor nanomaterials by integration onto piezoelectric actuators“, *Semicond. Sci. Technol.*, Bd. 33, Nr. 1, S. 013001, Jan. 2018, doi: 10.1088/1361-6641/aa9b53.
[24] X. Zhou *u. a.*, „Review on piezoelectric actuators: materials, classifications, applications, and recent trends“, *Front. Mech. Eng.*, Bd. 19, Nr. 1, S. 6, Feb. 2024, doi: 10.1007/s11465-023-0772-0.
[25] A. Laneve *u. a.*, „Wavevector-resolved polarization entanglement from radiative cascades“, *Nat. Commun.*, Bd. 16, Nr. 1, S. 6209, Juli 2025, doi: 10.1038/s41467-025-61460-3.
[26] J. M. Liu, „Simple technique for measurements of pulsed Gaussian-beam spot sizes“, *Opt. Lett.*, Bd. 7, Nr. 5, Art. Nr. 5, Mai 1982, doi: 10.1364/OL.7.000196.
[27] K. Bischoff, D. Mücke, G.-L. Roth, C. Esen, und R. Hellmann, „UV-Femtosecond-Laser Structuring of Cyclic Olefin Copolymer“, *Polymers*, Bd. 14, Nr. 14, S. 2962, Juli 2022, doi: 10.3390/polym14142962.
[28] Y. Zhao, Y.-L. Zhao, und L.-K. Wang, „Application of femtosecond laser micromachining in silicon carbide deep etching for fabricating sensitive diaphragm of high temperature pressure sensor“, *Sens. Actuators Phys.*, Bd. 309, S. 112017, Juli 2020, doi: 10.1016/j.sna.2020.112017.
[29] H. K. Toenshoff, A. Ostendorf, und T. Wagner, „Structuring silicon with femtosecond lasers“, gehalten auf der Photonics West 2001 - LASE, M. C. Gower, H. Helvajian, K. Sugioka, und J. J. Dubowski, Hrsg., San Jose, CA, Juni 2001, S. 88–97. doi: 10.1117/12.432500.
[30] C. Fornaroli, J. Holtkamp, und A. Gillner, „Dicing of Thin Si Wafers with a Picosecond Laser Ablation Process“, *Phys. Procedia*, Bd. 41, S. 603–609, 2013, doi: 10.1016/j.phpro.2013.03.122.
[31] C. Kalupka, D. Großmann, und M. Reininghaus, „Evolution of energy deposition during glass cutting with pulsed femtosecond laser radiation“, *Appl. Phys. A*, Bd. 123, Nr. 5, S. 376, Mai 2017, doi: 10.1007/s00339-017-1000-8.
[32] S. Stroj, W. Plank, und M. Muendlein, „Ultrashort-pulsed laser separation of glass-silicone-glass substrates: influence of material properties and laser parameters on dicing process and cutting edge geometry“, *Appl. Phys. A*, Bd. 127, Nr. 1, S. 7, Jan. 2021, doi: 10.1007/s00339-020-04192-z.
[33] M. B. Rota *u. a.*, „A source of entangled photons based on a cavity-enhanced and strain-tuned GaAs quantum dot“, *eLight*, Bd. 4, Nr. 1, S. 13, Dez. 2024, doi: 10.1186/s43593-024-00072-8.

[34] F. B. Basset *u. a.*, „Signatures of the Optical Stark Effect on Entangled Photon Pairs from Resonantly-Pumped Quantum Dots“, 2. August 2023, *arXiv*: arXiv:2212.07087. Zugegriffen: 9. November 2023. [Online]. Verfügbar unter: http://arxiv.org/abs/2212.07087
[35] B. U. Lehner *u. a.*, „Beyond the Four-Level Model: Dark and Hot States in Quantum Dots Degrade Photonic Entanglement“, *Nano Lett.*, Bd. 23, Nr. 4, S. 1409–1415, Feb. 2023, doi: 10.1021/acs.nanolett.2c04734.
[36] X. L. Zhang *u. a.*, „Optical bandgap and phase transition in relaxor ferroelectric Pb(Mg1⁄3Nb2⁄3)O3-*x*PbTiO3 single crystals: An inherent relationship“, *Appl. Phys. Lett.*, Bd. 103, Nr. 5, Juli 2013, doi: 10.1063/1.4816965.